\documentclass{appolb}
\usepackage{epsfig}
\usepackage[T1]{fontenc}
\usepackage{graphicx}
\usepackage{amsfonts}
\usepackage{subfigure}

\begin{document}

 \eqsec
\title{A model of loops in 2-D%
\thanks{Presented at the workshop Random Geometry Krakow 2003}}
\author{Aleix PRATS
  \address{Departament d'Estructura i Constituents de la Materia\\%
    Facultat de F\'\i sica, Universitat de Barcelona\\%
    Diagonal 647, 08028 Barcelona, Spain}}
\newcommand{\dif}{\mathrm{d}}
\newcommand{\rmspc}{\!\!\!\!\!\!\!}
\newcommand{\barint}{-\!\!\!\!\!\!\!\!\:\int}

\maketitle

\begin{abstract}
The gonihedric spin model was first introduced as the action for a
discretized tensionless string in a discretized embeding
space. Afterwards was found that there are interesting features on the
dynamical behavior of this model in 3 dimensions (as it was first
formulated) that make us think on glassy spin model without inherent
disorder. Extensive simulations have been carried out in the 3 dimensional
model. In the following I will report on a work composed of two
different but related parts. The first part is a numerical study
through Monte Carlo simulations of the dynamical properties of the 2
dimensional version of the model (\ie the loop model), which is much
simpler due to the fact that it has trivial thermodynamical
properties. The second part consists on an analytical approach of this
2 dimensional loop model coupled to gravity. We solve partially the
associated two-matrix model via a reduction to an equivalent one matrix
model and saddle point methods with the last one-matrix model.
\end{abstract}

\section{An statistical model of loops in 2D}
The gonihedric spin model was first introduced by Savvidy  in
relation to a discretized model for a tensionless string theory
\cite{Amba-sukiasian}, but very soon the spin model gained interest by
itself. Also 
its extension to a self-interacting surfaces ($\kappa\ne 0$) showed a
very rich family of models with different kind of critical points and
interesting dynamical properties
\cite{Amba-sukiasian}\cite{savvidy}\cite{kuts}.
Extensive numerical and theoretical work appeared
\cite{baig}\cite{esbaijo}\cite{jonmal}\cite{eslipjo} and some 
interest about the glassiness of the 3-dimensional gonihedric model
arised \cite{swift}\cite{lip}\cite{lipjo}\cite{lipjo_}\cite{DEJP}. This is precisely the
aspect of the model 
that has motivated us to study the 2-dimensional version of the
model. Its trivial thermodynamics motivates us to investigate whether
this model also has glassy behavior or not. This would provide a toy
model for glassy phenomena without inherent disorder on the couplings.

So the spin model we are going to investigate is related to a model of
loops in 2 dimensions. We are going to compare this model and the
results we obtained from analogous simulations in a 3 dimensional
version of this spin model \cite{DEJP}. But before we go into the
analysis of the model let's define it in terms of spin variables.

Consider the following Hamiltonian

\begin{equation}
\mathcal{H}_{gonih}^{\textit{\tiny 2D}}=
-\kappa\sum_{<i,j>}\sigma_i\sigma_j 
+\frac{\kappa}{2}\sum_{\ll i,j\gg}\sigma_i\sigma_j
+\frac{1-\kappa}{2}\sum_{[i,j,k,l]}\sigma_i\sigma_j\sigma_k\sigma_l
\nonumber 
\end{equation}
in a two dimensional lattice\footnote{The 3 dimensional counterpart
  has slightly modified couplings. the explicit form is \\ 
$$\mathcal{H}_{gonih}^{\textit{\tiny
  2D}}=-2\kappa\sum_{<i,j>}\sigma_i\sigma_j +\frac{\kappa}{2}\sum_{\ll
  i,j\gg}\sigma_i\sigma_j +\frac{1-\kappa}{2}\sum_{[i,j,k,l]} $$ \\ 
  The reason is that the number of neighbors changes from 3D to 2D, thus the
  couplings has to change too as we will argue.}, where $<i,j>$ means
  sum over nearest neighbors, $\ll i,j\gg$ means sum over next to
  nearest neighbors, and $[i,j,k,l]$ means over spins forming
  plaquettes on the lattice. 

This  Hamiltonian has some odd characteristics. The most important of
  all is that the space of symmetric vacua is extremely large, in
  fact it is exponentially large with the dimension of the lattice
  $L$. In particular, the simultaneous flip of all the spins that
  belong to any set of non-crossing lines leaves the energy of
  the ferromagnetic ground state unchanged\footnote{In the 3
  dimensional case this symmetry is generated by the flip of planes
  rather than lines.} \cite{sav}. This symmetry is even larger in the
  $\kappa=0$ case where the lines can cross each other. This provides
  a very special landscape for the energy function of our model that
  in its 3 dimensional version makes the system exhibit a very clear
  glassy behavior associated to a thermodynamical phase transition
  \cite{lipjo}. This is precisely the aim of this work: to determine
  whether or not the same kind of behavior can be found in 2
  dimensions given that the 2D model has no thermodynamical phase
  transition. 

Let us first see the relation between this model and the loop model we
announced. If we look at the energy of a given configuration we can see
that due to the precise fine tuning of the couplings all the energy
is concentrated at the bending points of the loop (surface in this 3D case)
that is the boundary between the two different phases of the system
(i.e. between plus and minus spins), and that wherever there is a
crossing of this surface with itself (or with another loop) there is
another concentration of extra energy. So at the end of the day we can
write the energy of the spin model (or the loop model) as follows
\begin{equation} 
E=n_2+4\kappa n_4\nonumber
\end{equation}
where $n_2$ is the number of bending points and $n_4$ the number of
self-crossing points of the loop that separates the plus and minus
spins regions. 

\begin{figure}[!ht]
\begin{center}
\subfigure[]{\includegraphics[scale=0.23]{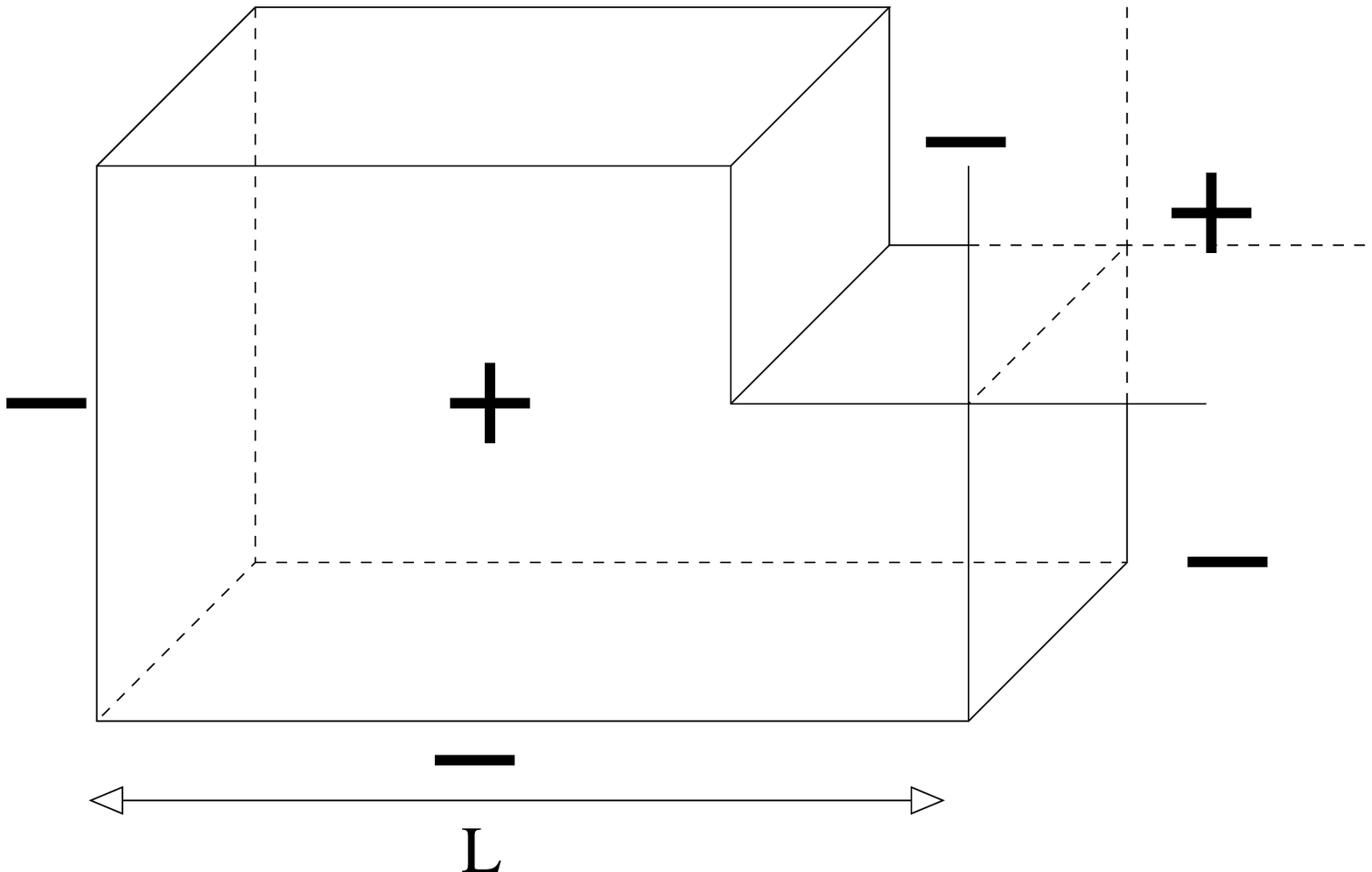}}
\subfigure[]{\includegraphics[scale=0.23]{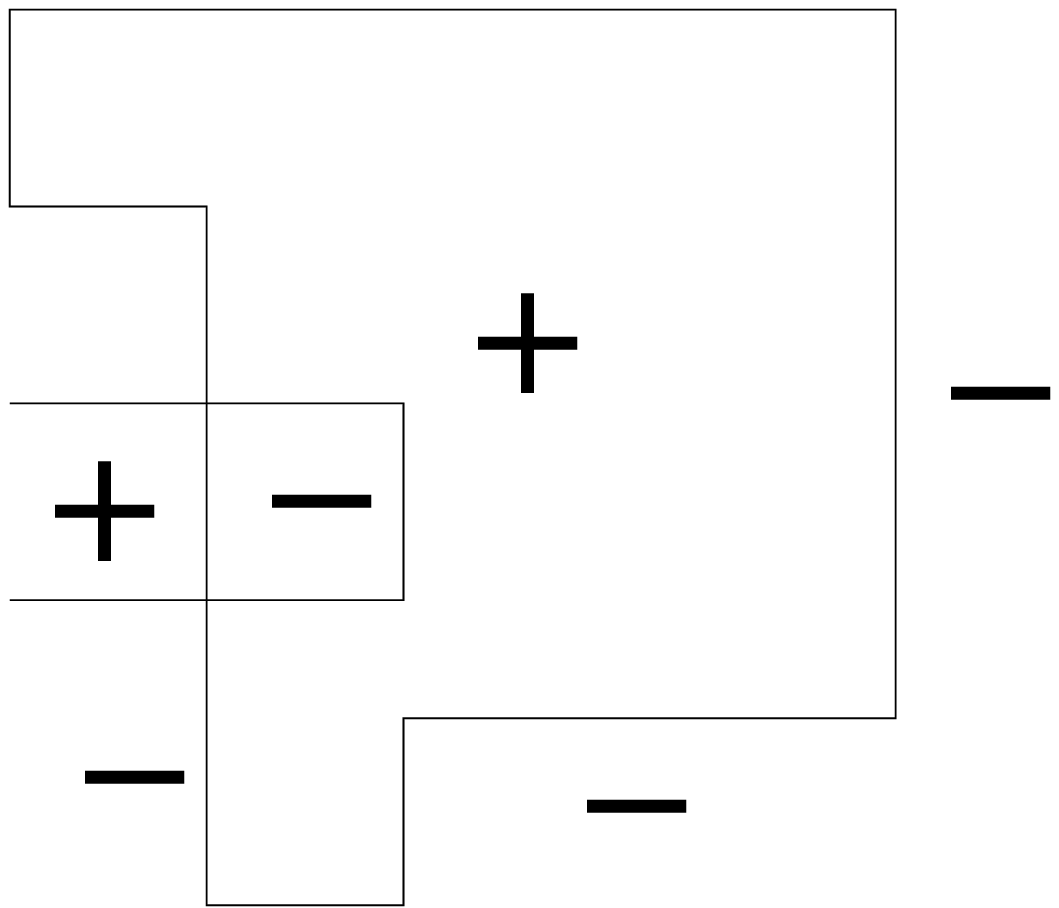}}
\end{center}
\caption{Examples of configurations of the system in 3D (a) and in 2D
  (b). The energy is concentrated on the bending and crossing  points
  (2D) and lines (3D).}  
\label{2-3-Dexample}
\end{figure}

This is exactly the same that happens in the 3 dimensional version
of the model. In fact the couplings are precisely chosen to exhibit
these features. But all this has very different consequences for
2 dimensions or 3 dimensions. The main difference between this 2D
loop model and the corresponding 3D surface model is that the action
of the surface in 3D is proportional to the linear extent of the surface (see
fig.\ref{2-3-Dexample}a) and the roughness of it, while the action for
the loop is not depending on how big it is but on how many times it
bends (see fig.\ref{2-3-Dexample}b).
Thus in 2 dimensions the energy of the loop do not depend on its
size but on the shape of it (although the energy barriers do
eventually depend on the size of the loop).

\subsection{Thermodynamical behavior}

\begin{figure}[!ht]
\begin{center}
\subfigure[]{\includegraphics[scale=0.23,angle=-90]{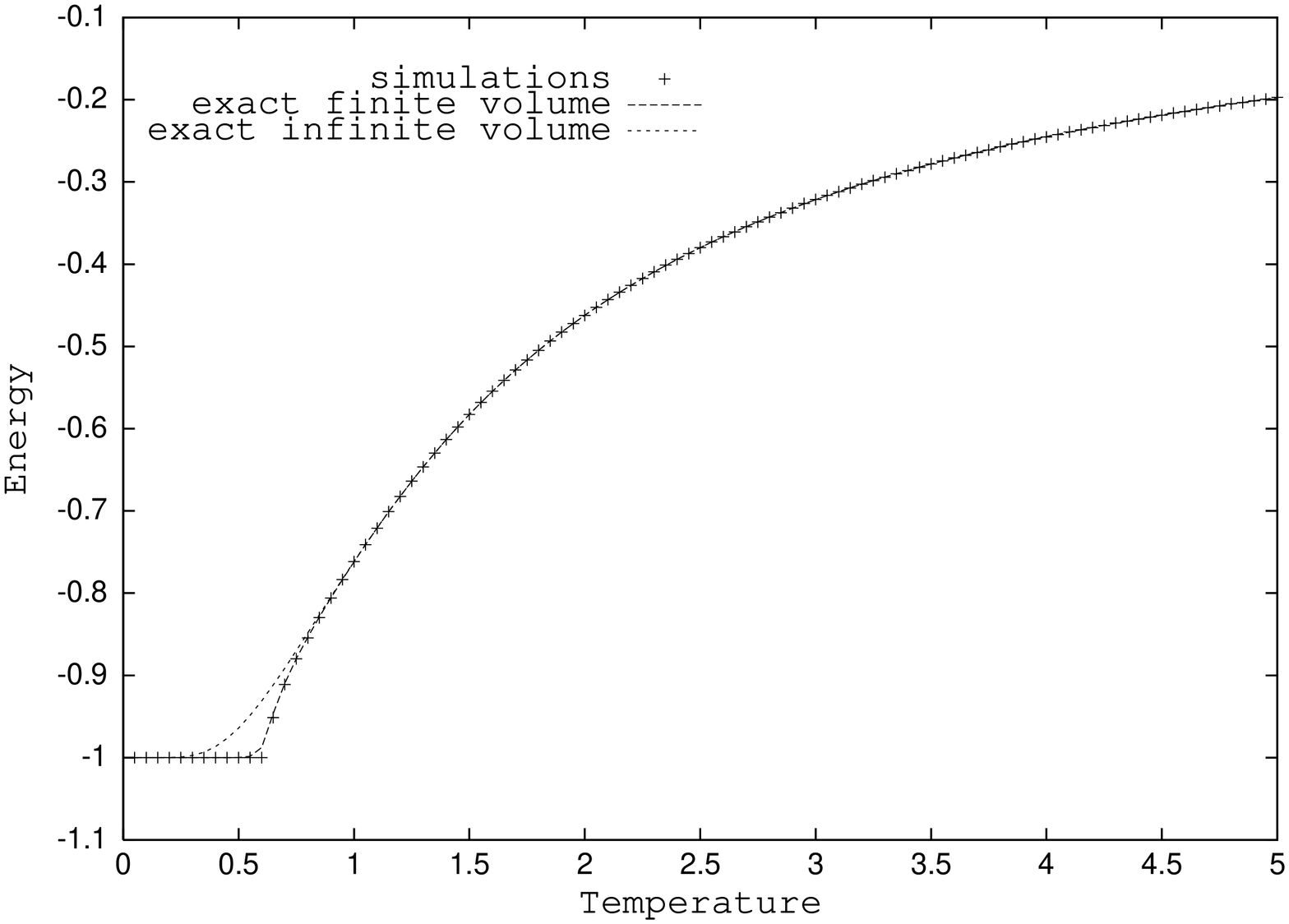}}
\subfigure[]{\includegraphics[scale=0.23,angle=-90]{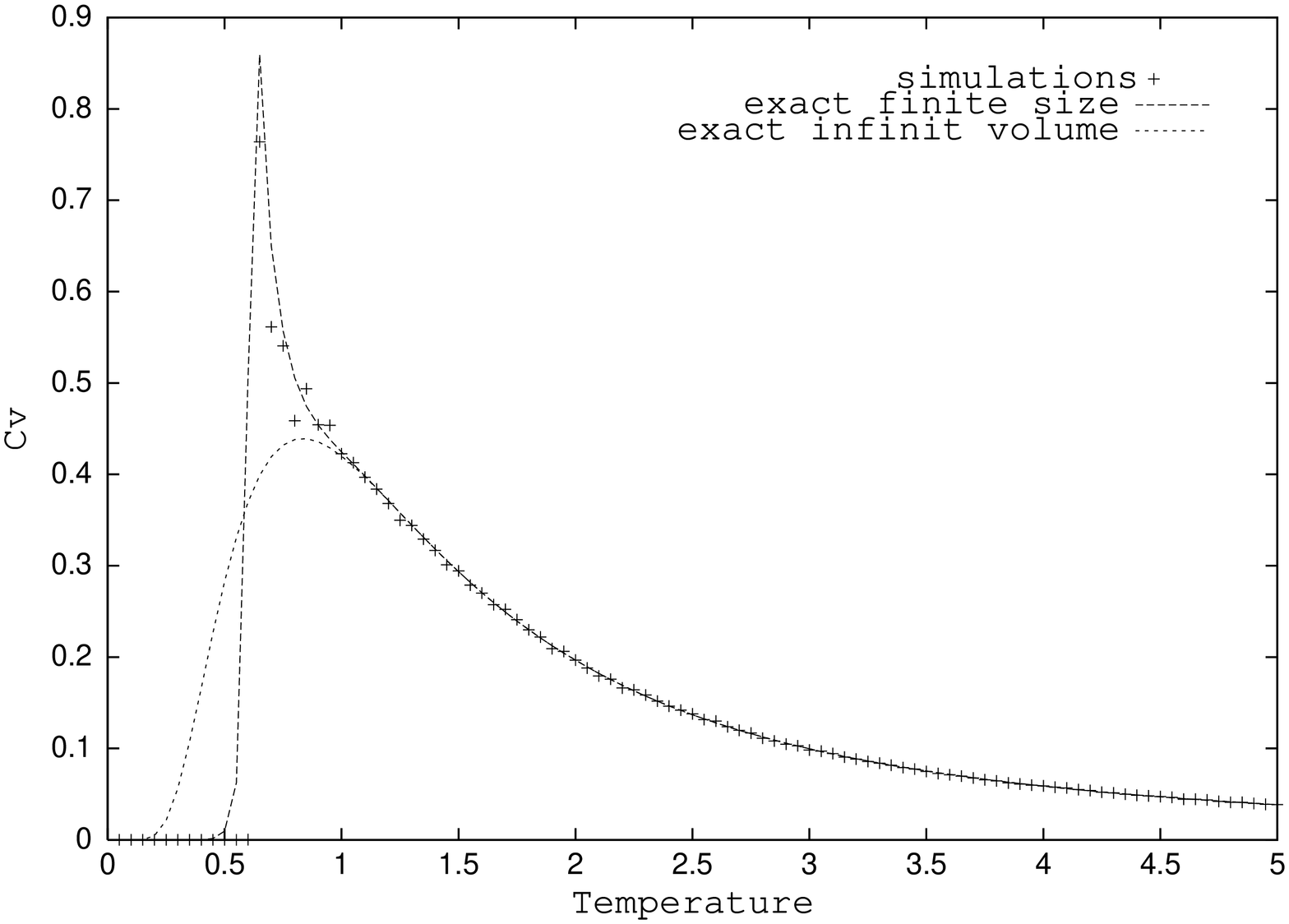}}
\end{center}
\caption{(a) Energy function and (b) specific heat of the system for
  $\kappa=0$. The exact function at infinite volume, at finite volume,
  and the Monte-Carlo simulations are plotted}
\label{EneK0}
\end{figure}

\begin{figure}[!ht]
\begin{center}
\subfigure[]{\includegraphics[scale=0.23,angle=-90]{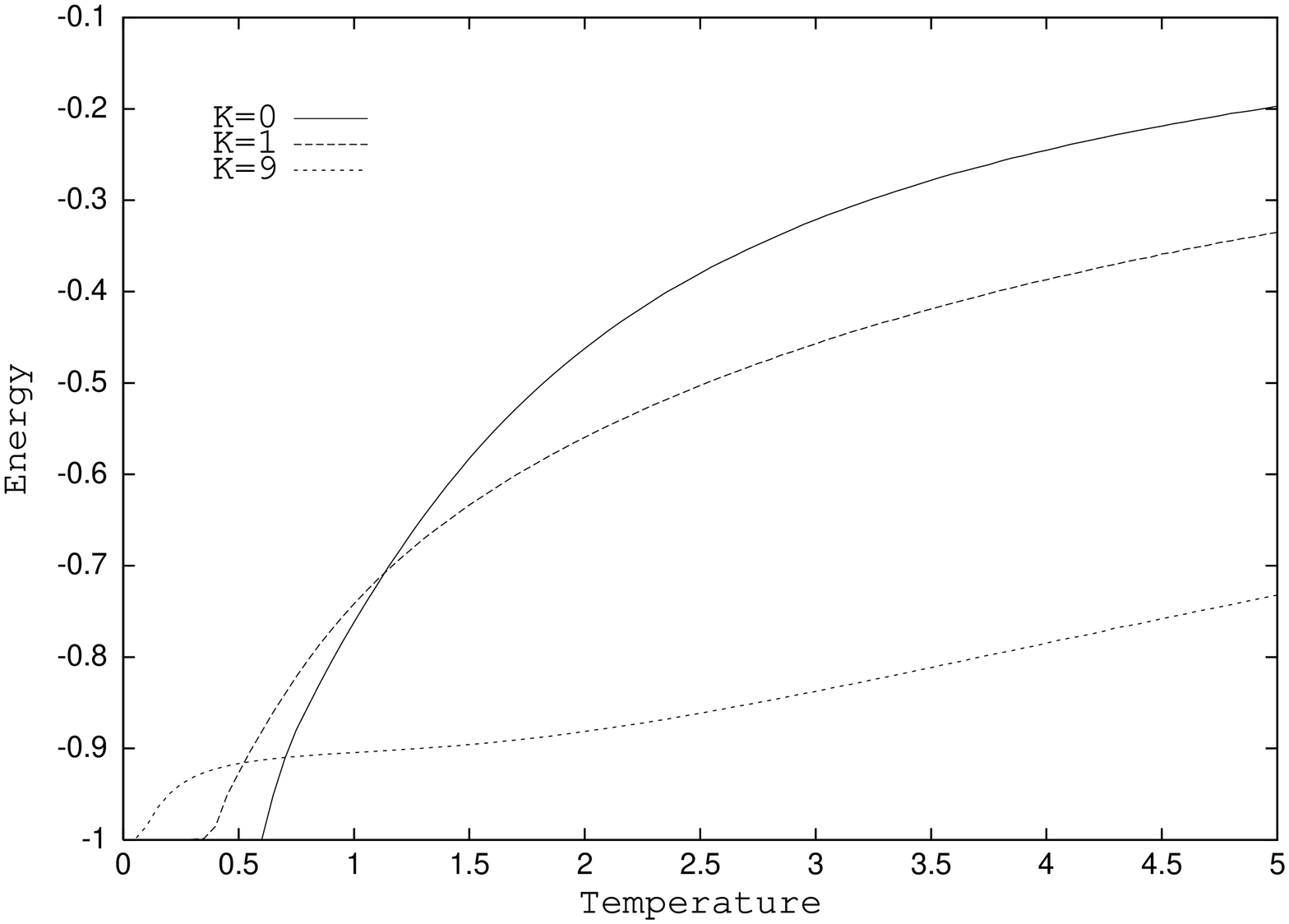}}
\subfigure[]{\includegraphics[scale=0.23,angle=-90]{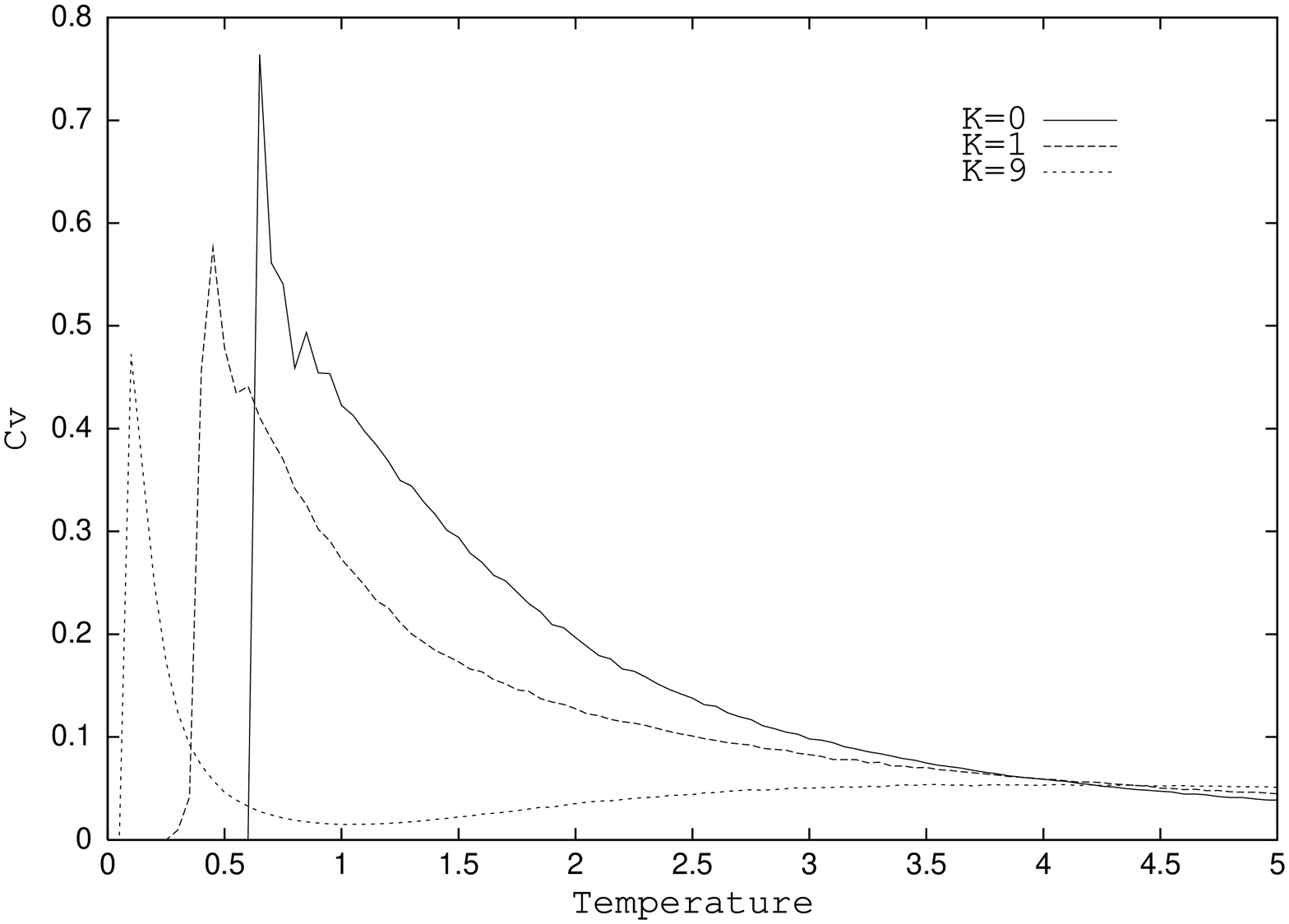}}
\end{center}
\caption{(a) Energy function and (b) specific heat of the system for
  $\kappa=0$. The exact function at infinite volume and the
  Monte-Carlo simulations are plotted} 
\label{EneKno0}
\end{figure}

Let's now take a look at the thermodynamical properties of this spin
model taking as a reference its 3 dimensional counterpart
\cite{esbaijo}\cite{DEJP}. Let's begin with the special case of
$\kappa=0 $ that is exactly solvable in infinite volume and reducible
to an easy-computable sum for finite volume.  The exact solution for
the model with $\kappa =0$ shows us that there is no phase transition
at finite temperature. If we take a look at fig.\ref{EneK0} we will
see the infinite volume energy function and susceptibility compared to
the numerical results of simulations and to the exact finite volume
calculation. All the discordances between simulations and the infinite
volume calculations are due to finite volume effects as we can see
comparing the simulations with the exact finite volume calculations.
For the other cases with $\kappa\ne 0$ there is no infinite volume
exact solution nor easy-computable finite-volume expression but the
simulations performed do not show great differences with the $\kappa
=0$ case (see fig.\ref{EneKno0}). The only remarkable difference is the
appearance of a second structure for sufficiently large $\kappa$. This
second structure can be interpreted as the appearance of a new energy
level for the plaquette variables. This has been studied to see
whether it evolves into a peak at large volumes, but no volume
dependence of this structure has been found, so there is no evidence
of phase transition.  

\begin{figure}[!ht]
\begin{center}
\includegraphics[scale=0.3,angle=-90]{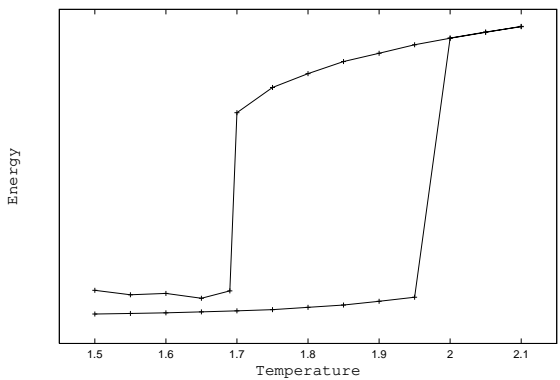}
\end{center}
\caption{Energy versus temperature. The lower branch is produced
  heating the ferromagnetic ground state. The higher branch
  corresponds to sudden quenches at each temperatures from disordered
  configurations.}
\label{3DPhaseSpace}
\end{figure}

On the other hand the same model in three dimensions exhibits a quite
complex phase space. For $\kappa =0$ there is a critical temperature
$T_c$ where the system changes from ordered to disordered phase
through a first order phase transition, and a second temperature $T_g$
that is between two different dynamical phases: a glassy phase and a
supercooled phase (see fig.\ref{3DPhaseSpace})
\cite{lip}\cite{lipjo_}\cite{kutsanis}. Increasing $\kappa$ we find
from certain value on that this second temperature $T_g$ either is
very close to the thermodynamical temperature $T_c$ or they coincide,
and that this thermodynamical phase transition changes from first to
second order.

Thus we have seen that this models from 3 dimensions to 2 dimensions
changes a lot its thermodynamical behavior. From having first/second
order phase transitions (depending on the value of $\kappa$) on 3D to
trivial thermodynamics without any phase transition on 2D. But we also
want to know whether there is a great difference or not in they dynamical
properties. In particular we want to know if the slow dynamics and the
glassy behavior remains on the 2 dimensional model.  

\subsection{Dynamical behavior}

We will move now to the dynamical properties of
the model. We shall report here only the $\kappa=0$ case\footnote{Due
  that we have not established yet without ambiguities whether this
  case posses or not glassy behavior we are not going to consider the
  $\kappa\ne 0$ case.}.

\begin{figure}[!ht]
\begin{center}
\includegraphics[scale=0.3,angle=-90]{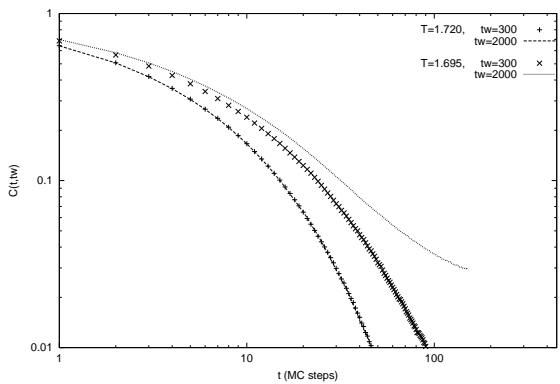}
\end{center}
\caption{curves of autocorrelation of the spin variables at both sides
of $T_g$ on the 3 dimensional gonihedric model} 
\label{Cdtslow3D}
\end{figure}

\begin{figure}[!ht]
\begin{center}
\includegraphics[scale=0.3,angle=-90]{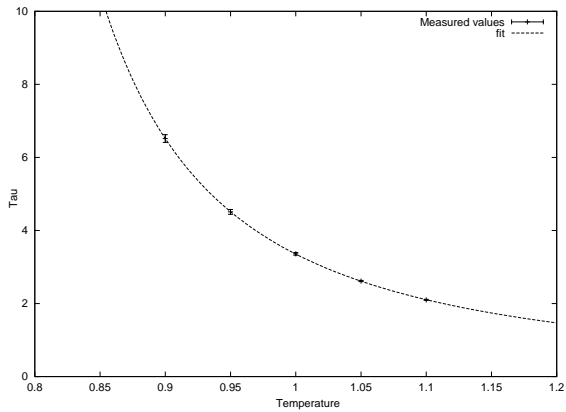}
\end{center}
\caption{Autocorrelation time versus Temperature at temperatures above
0.9}
\label{tau2D}
\end{figure}

\begin{figure}[!ht]
\begin{center}
{\includegraphics[scale=0.3,angle=-90]{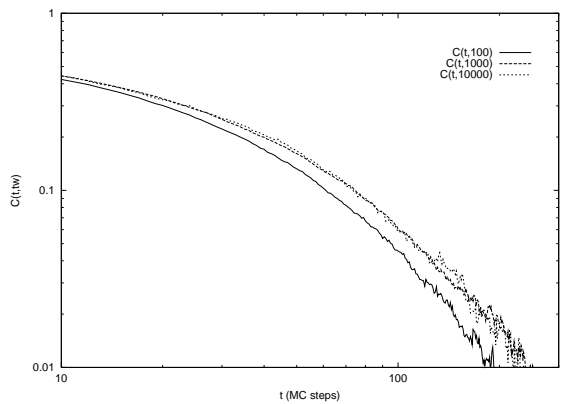}}
\end{center}
\caption{Evolution of the autocorrelation function with the waiting
  time $t_w$.}
\label{Thermal.1}
\end{figure}

To study the dynamics of this system we will consider a two-time
correlator of local observables \cite{mezard}
\begin{equation}
C(t,t_w)=\sum_i e_i(t_w) e_i(t_w+t)\nonumber
\end{equation}
where the sum runs over all the sites in the lattice, and the variable
$e_i(t)$ is the energy\footnote{We could have used other kind of
  observables like the spin variables or the energy per plaquette, but
  they have the same behavior for our purposes.} of the site $i$ at the
time $t$. This object, in equilibrium, should be independent of $t_w$,
but as we
\begin{figure}[!ht]
\begin{center}
\subfigure[]{\includegraphics[scale=0.23,angle=-90]{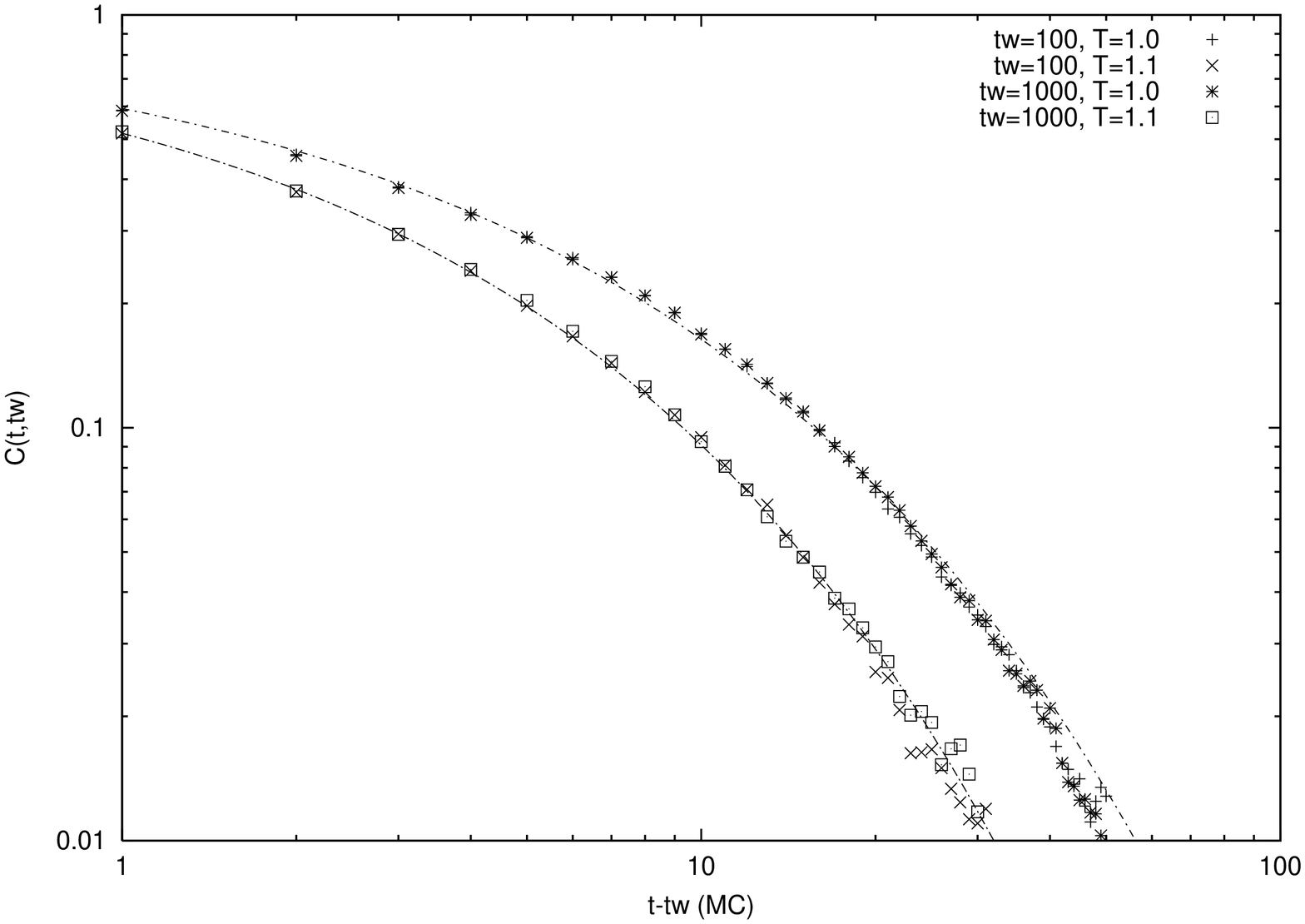}}
\subfigure[]{\includegraphics[scale=0.23,angle=-90]{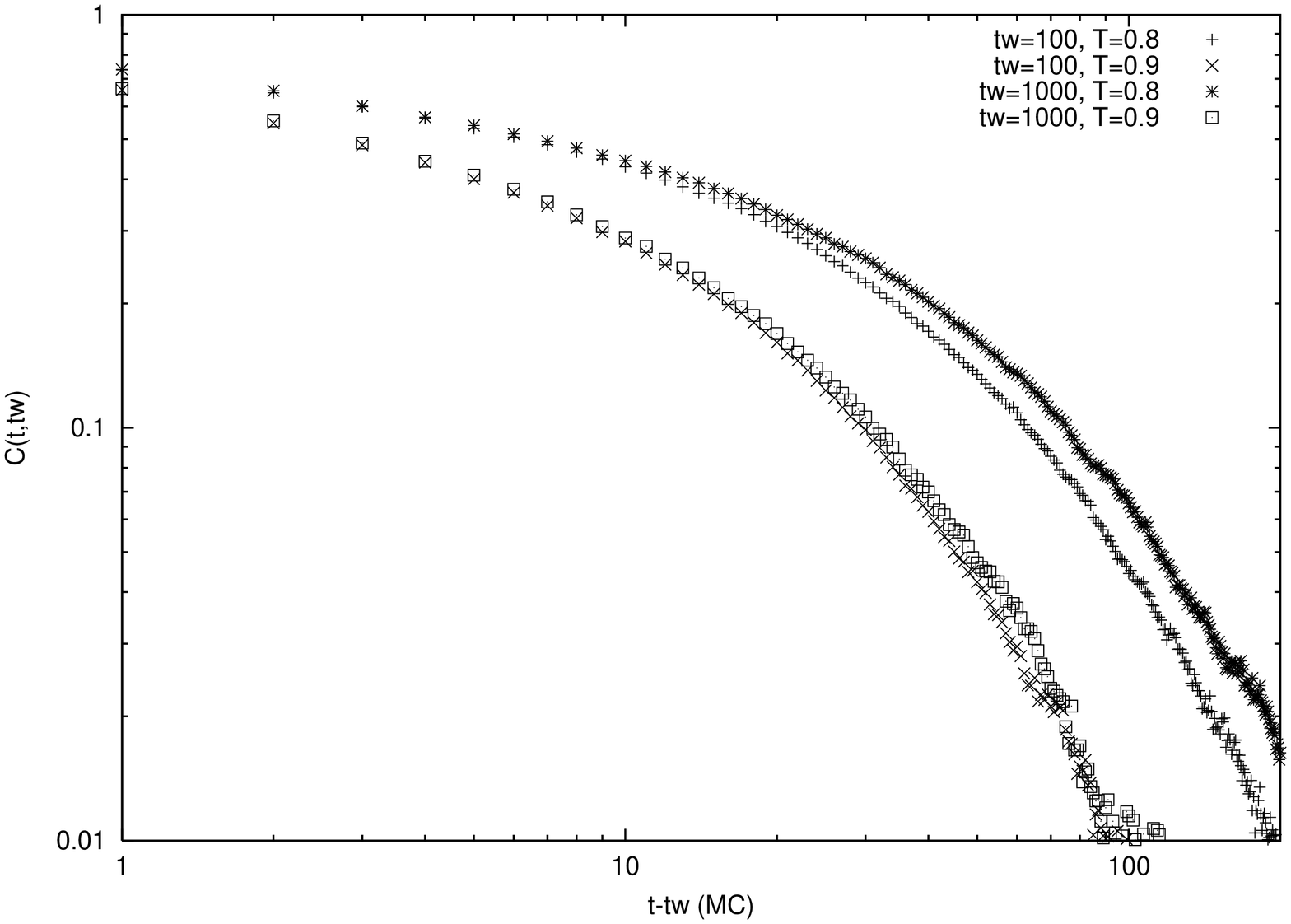}}
\end{center}
\caption{Autocorrelation functions for different temperatures and
  different waiting times: (a) curves are independent at $t_w$ (b) some
  dependence in $t_w$ appears} 
\label{Cdtslow2D}
\end{figure}
can see in fig.\ref{Cdtslow2D} below some temperature this function
happens to depend on the waiting time $t_w$, such observation is
identical to the one made for the 3 dimensional case where an even larger
dependence of this autocorrelation function on the waiting time
appeared below $T_g$ (see fig.\ref{Cdtslow3D}). 
This is not per se a clear evidence for glassy behavior of the model yet,
so we will continue with the program we followed with the
3 dimensional model. Thus let's fit the curves that look
$t_w$-independent (i.e. the curves that should be above the
hypothetical glassy transition temperature$T_g$). The fitting function
will be of the form 
\begin{equation}
Ae^{(\frac{t}{\tau})^b}
\label{streched}
\end{equation}
which is an stretched exponential. As you can see in
fig.\ref{Cdtslow2D}a the agreement between the fit and the simulations
is rather good\footnote{Lines are fits, points are simulation
  measurements. We looked for the consistency of the
fits by checking that the true value of the parameter $A$, i.e. $A=1$
(which we know by construction of the correlation function) were within the
interval of confidence of the fitted value}. Now we can plot the
fitted values of $\tau$ against temperature, and we see that it
increases as we lower the temperature as if it liked to diverge at
some point. If we fit this points using a function of the
form 
\begin{equation}
\frac{\tau_o}{(T-T_*)^c}\nonumber
\end{equation}
as we did in 3 dimensions, we'll find a good parameterization of the
divergence (see fig.\ref{tau2D}) although there is still something
that is not completely compatible with this fit.  
The problem with this fit is that the fitted ``critical'' temperature
$T_*$ is not consistent with the point where the autocorrelation
function began to depend on the waiting time, as it should. This is
why we should analyze more carefully this autocorrelation function.
Looking more carefully at this autocorrelation function we can see
that the dependence in the waiting time $t_w$ disappears as we increase
it. This means that the thermalization of these two time
functions is extremely slow, but that could be non-glassy-like. In
fig.\ref{Thermal.1} we can see this convergence of the autocorrelation
functions as we increase the waiting time. 

\begin{figure}[!ht]
\begin{center}
\subfigure[]{\includegraphics[scale=0.23,angle=-90]{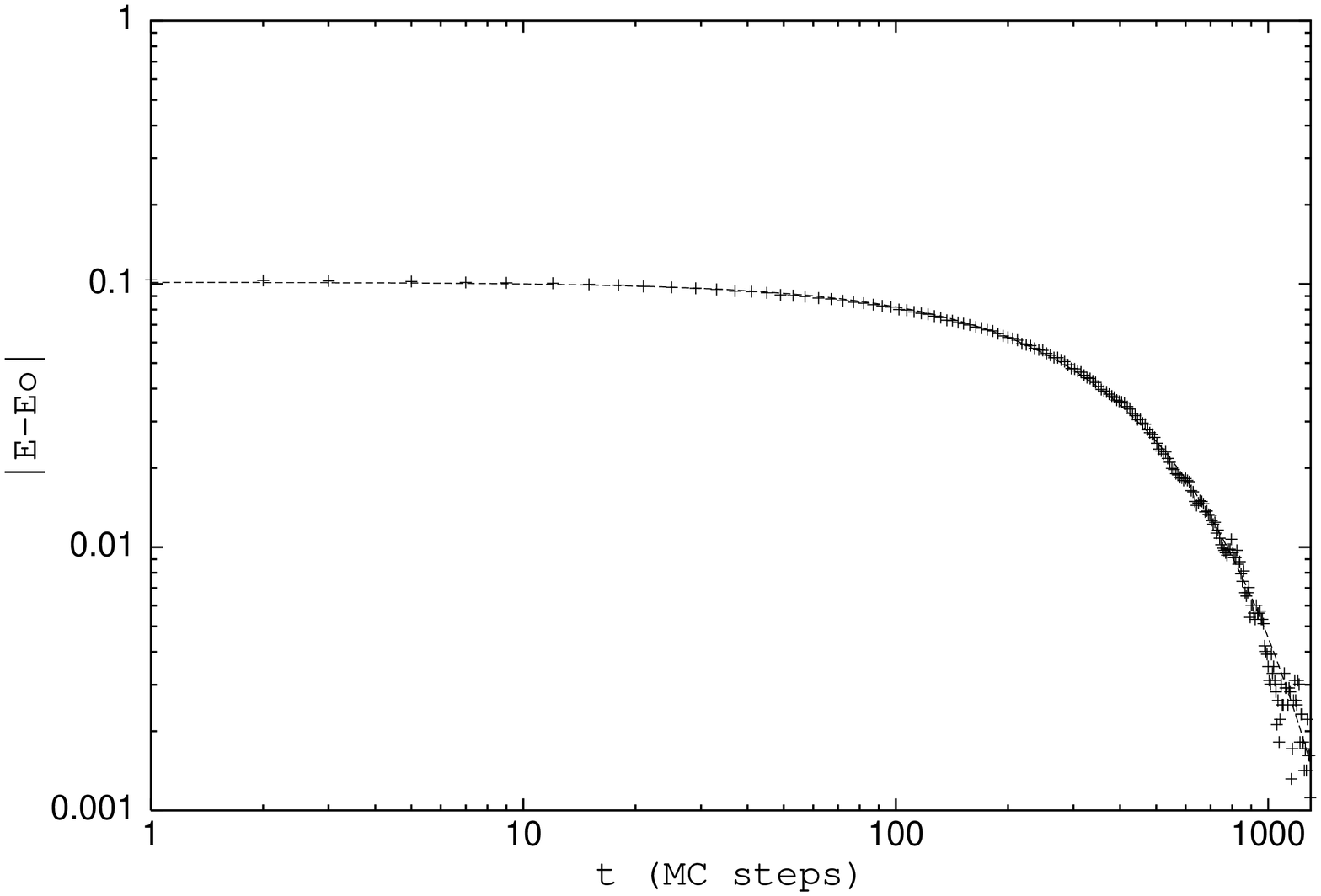}}
\subfigure[]{\includegraphics[scale=0.23,angle=-90]{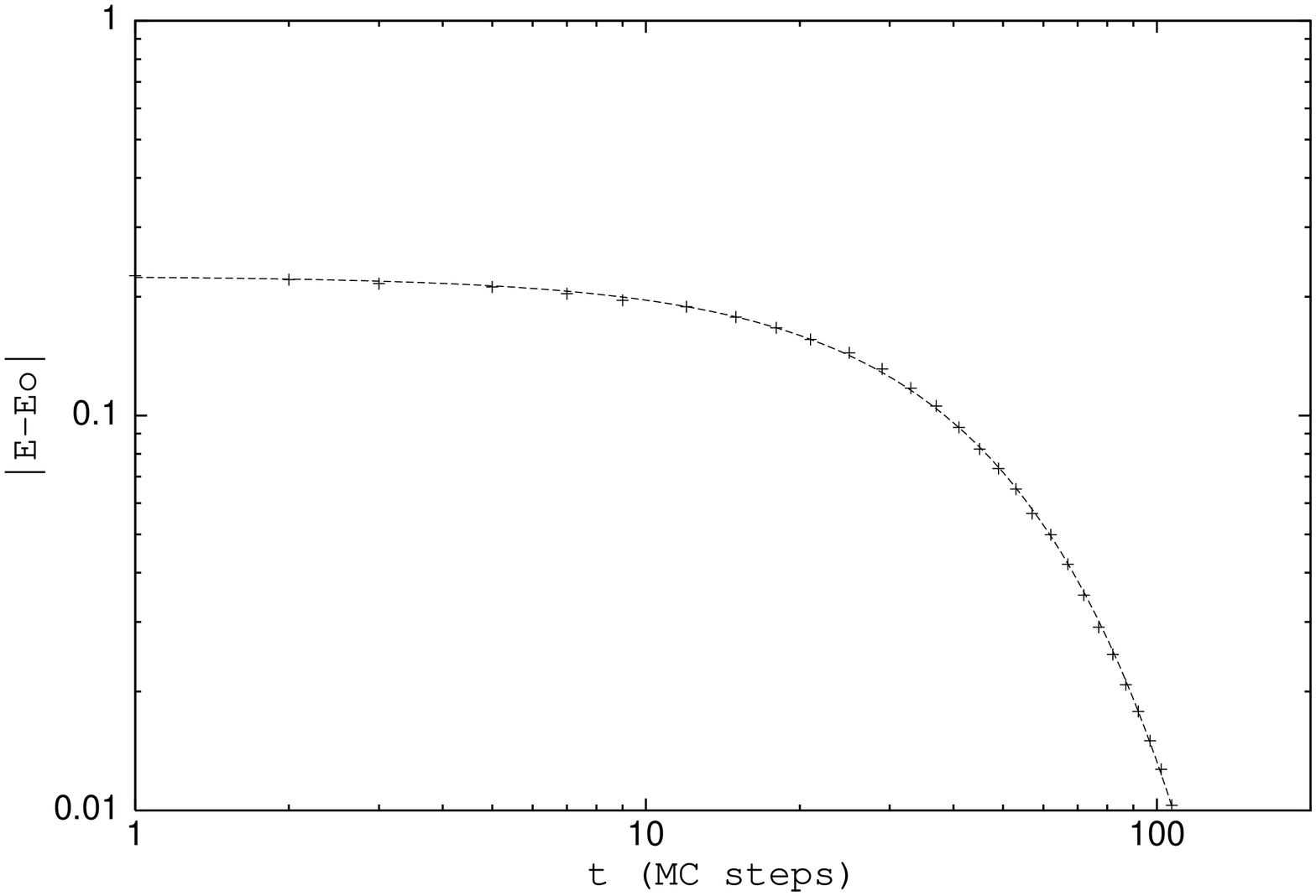}}
\end{center}
\caption{The decay of a prepared initial configuration to the equilibrium}
\label{2Ddecay}
\end{figure}

To explore better the dynamics behavior of this model we can perform other
type of experiments in our system. For example we can prepare our system in
a specific configuration, for example a lattice with two different
coexisting vacua, one inside the other (one possibility could be
layer-like vacua inside ferromagnetic one) and look at the decay
process of the system to the equilibrium. 
\begin{figure}[!ht]
\begin{center}
\subfigure[]{\includegraphics[scale=0.23,angle=-90]{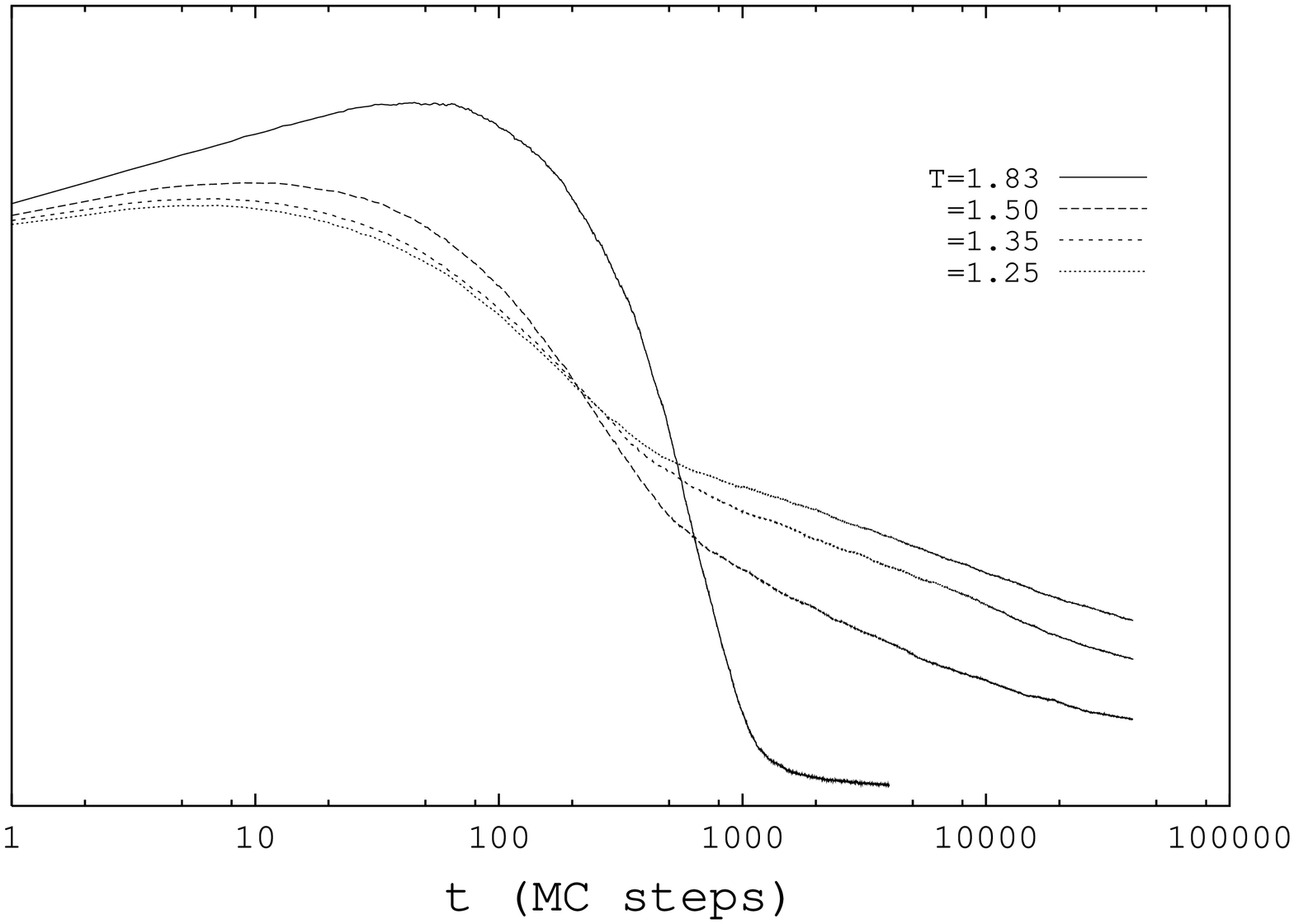}}
\subfigure[]{\includegraphics[scale=0.23,angle=-90]{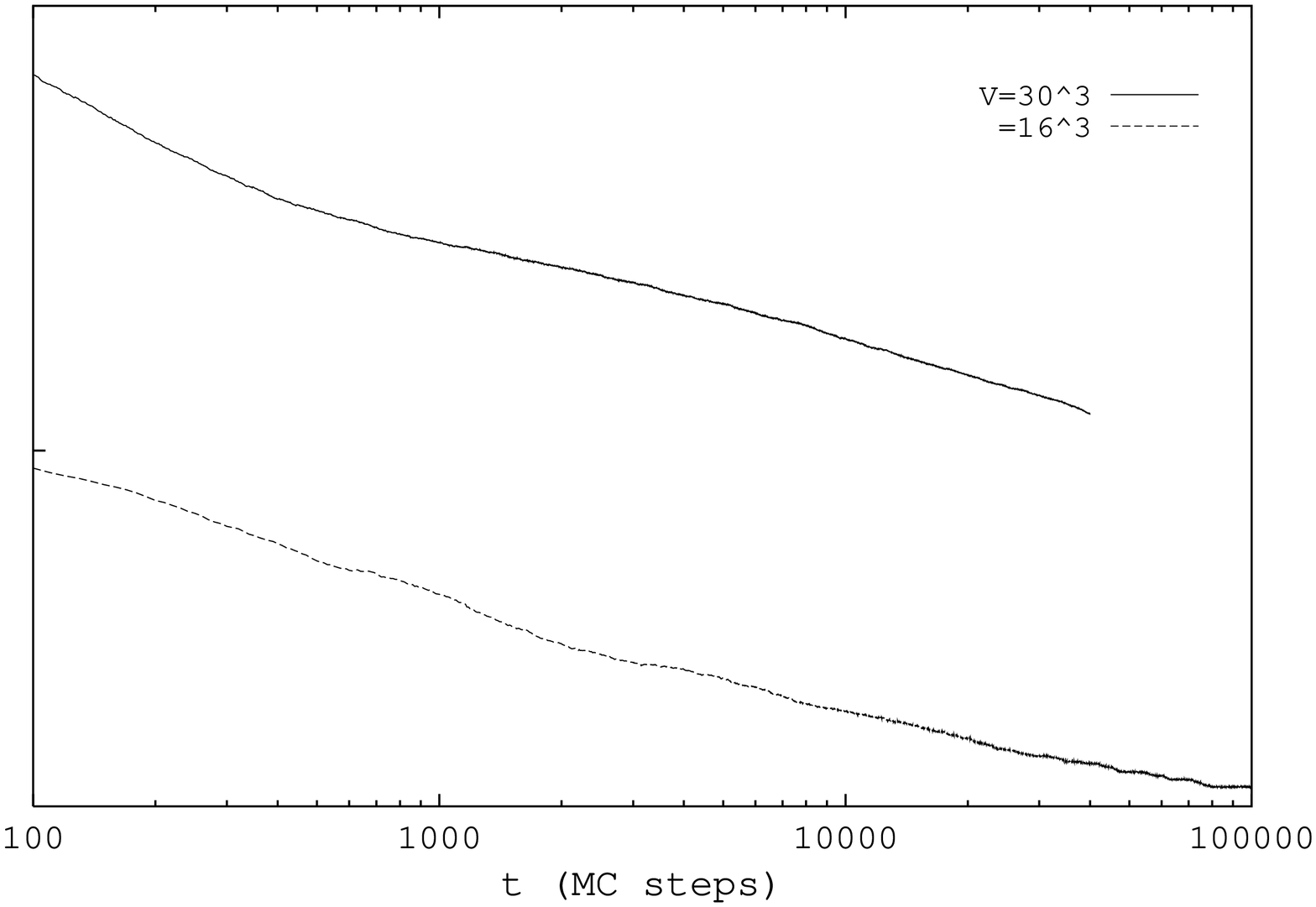}}
\end{center}
\caption{In the 3D gonihedric model: (a) Evolution of some order
  parameter starting from a perturbed vacua as an initial
  configuration. We can see two kind of behaviors. In (b) we see that
  the low temperature behavior is logarithmic. Lines are evolutions
  for different initial volumes of the perturbations at a temperature
  deep inside low temperature region.}   
\label{3Ddecay}
\end{figure}
The problem with this tests is that as there is no ordered phase in
this model, we cannot prepare the system in an initial configuration
composed by two different coexisting vacuas and pretend that they are
more disordered than the equilibrium-like configurations at the
temperatures we are examining, as it is in the 3 dimensional case
(fig.\ref{3Ddecay} shows the 3D decay of a perturbed vacua to the
unperturbed one. It is easy to see the different behaviors in terms of
temperature). in spite of this we performed those simulations and found
that the decay behaved in the same way for all the range of
temperatures (fig.\ref{2Ddecay}a and fig.\ref{2Ddecay}b are two
examples of this decays at both sides of the hypothetical $T_g$). The
magnitude we used to study this decay in two dimensions is the energy
difference with the equilibrium. If we look at the value of
the fitted exponent $c$ of the function (\ref{streched}) that we have
also used in this experiments, we can see that they are really close
to one, and this suggests that the behavior may not be glassy but
`usual' albeit rather slow exponential decay.

\begin{figure}[!ht]
\begin{center}
\includegraphics[scale=0.30,angle=-90]{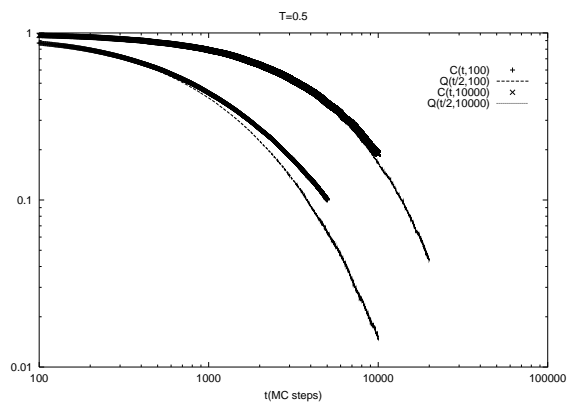}
\end{center}
\caption{The convergence to the scaling behavior is clear in this
  plot where $t_w$ has been increased by one order of magnitude.}
\label{convergeCdQ}
\end{figure}

In the analysis of the glassiness of the system we can introduce a new
observable; the $Q$ parameter \cite{mezard}. This parameter is defined in the
following way: after evolving a single system $t_w$ Monte Carlo steps,
we make two copies of the system and let them evolve
independently, then the $Q$ of a local observable is the overlap of
this observable between the two independent copies of the system. In
our case we will use the local spin variables 
\begin{equation}
Q=\sum_i \sigma_i^a(t_w+t)\sigma_i^b(t_w+t)
\end{equation}
Then, using this $Q$ parameter and the time overlap of the same local
magnitude, 
\begin{equation}
C_{spin}(t,t_w)=\sum_i \sigma_i(t_w) \sigma_i(t_w+t)\nonumber
\end{equation}
\begin{figure}[!ht]
\begin{center}
\subfigure[]{\includegraphics[scale=0.23,angle=-90]{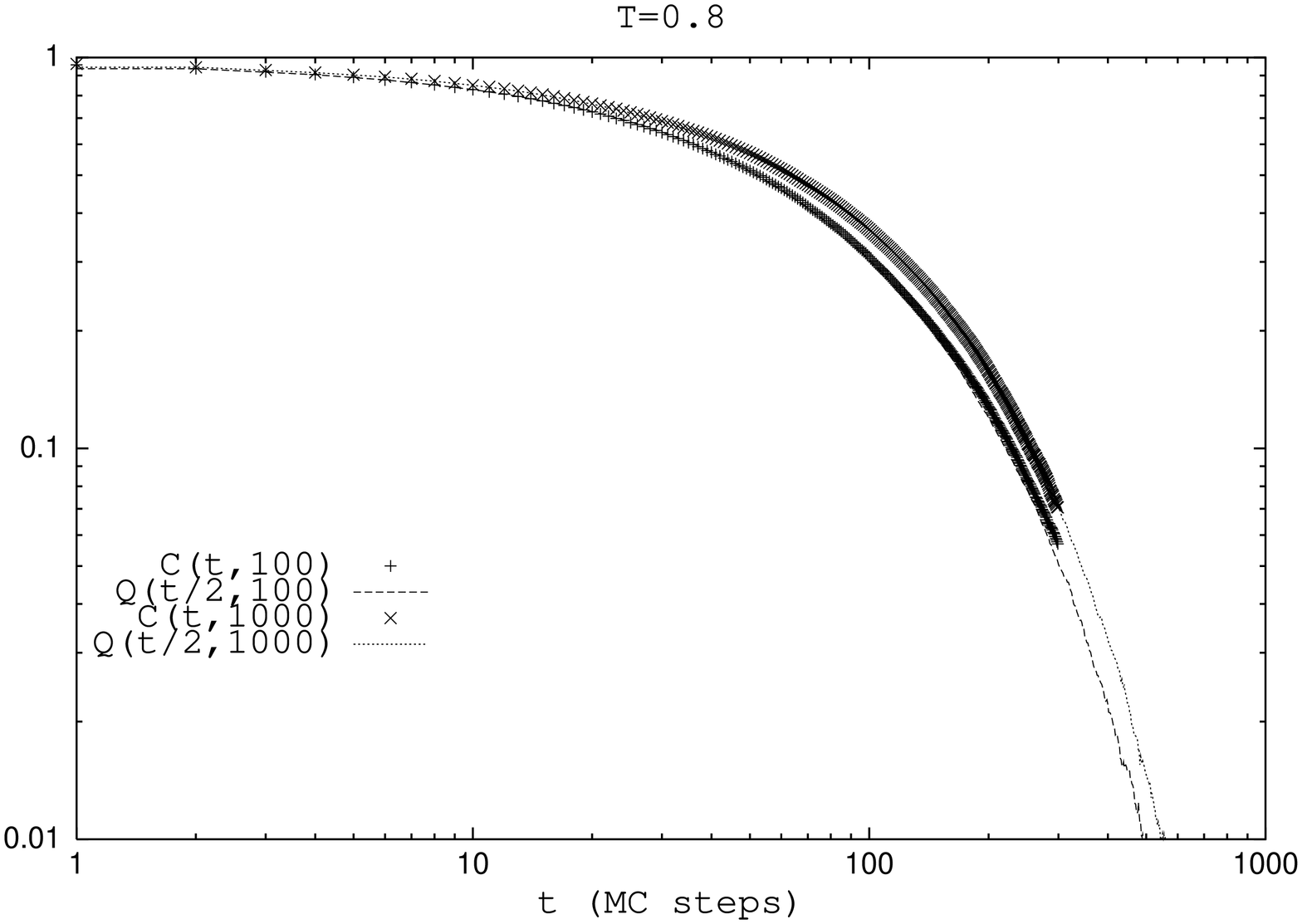}}
\subfigure[]{\includegraphics[scale=0.23,angle=-90]{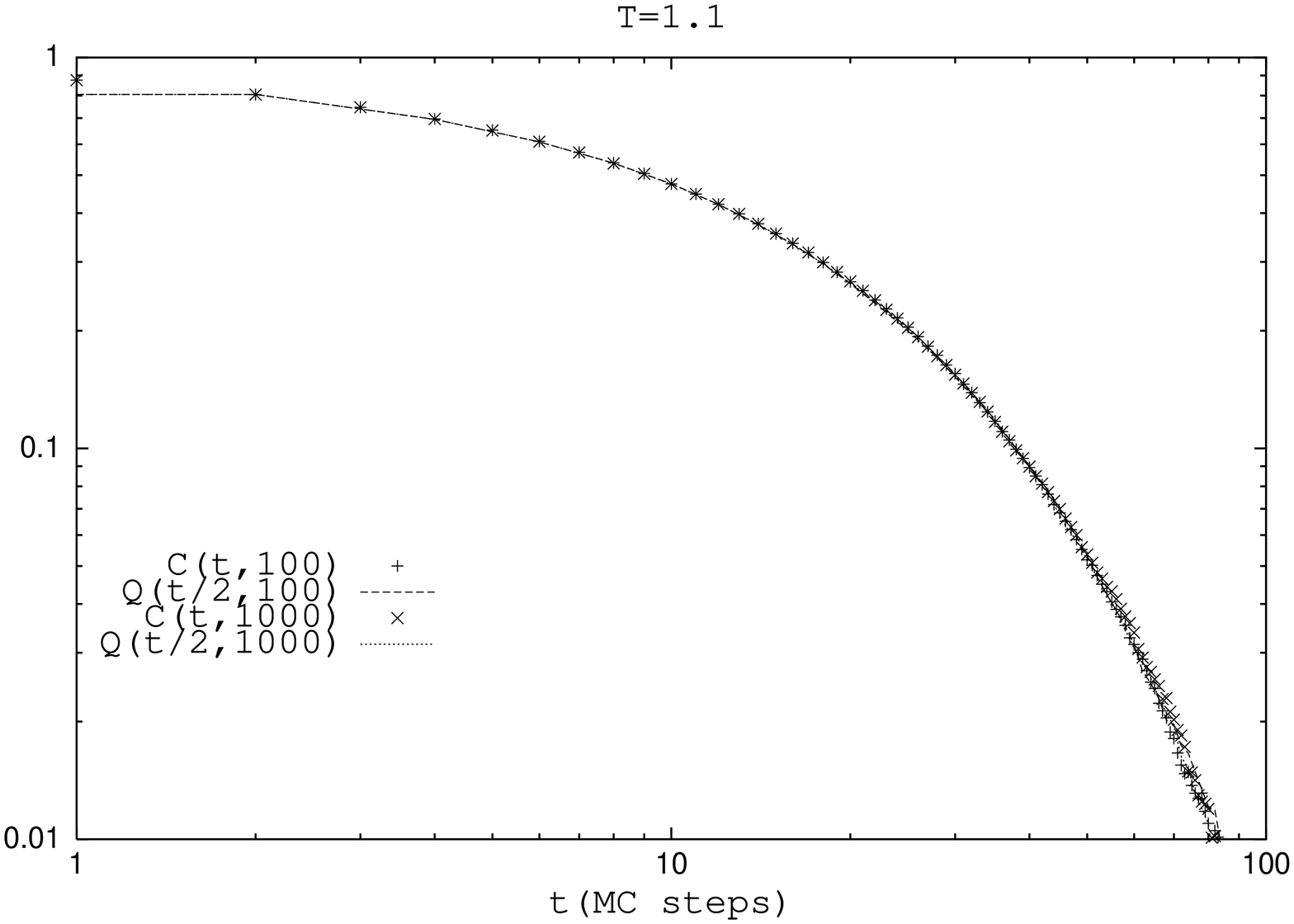}}
\subfigure[]{\includegraphics[scale=0.23,angle=-90]{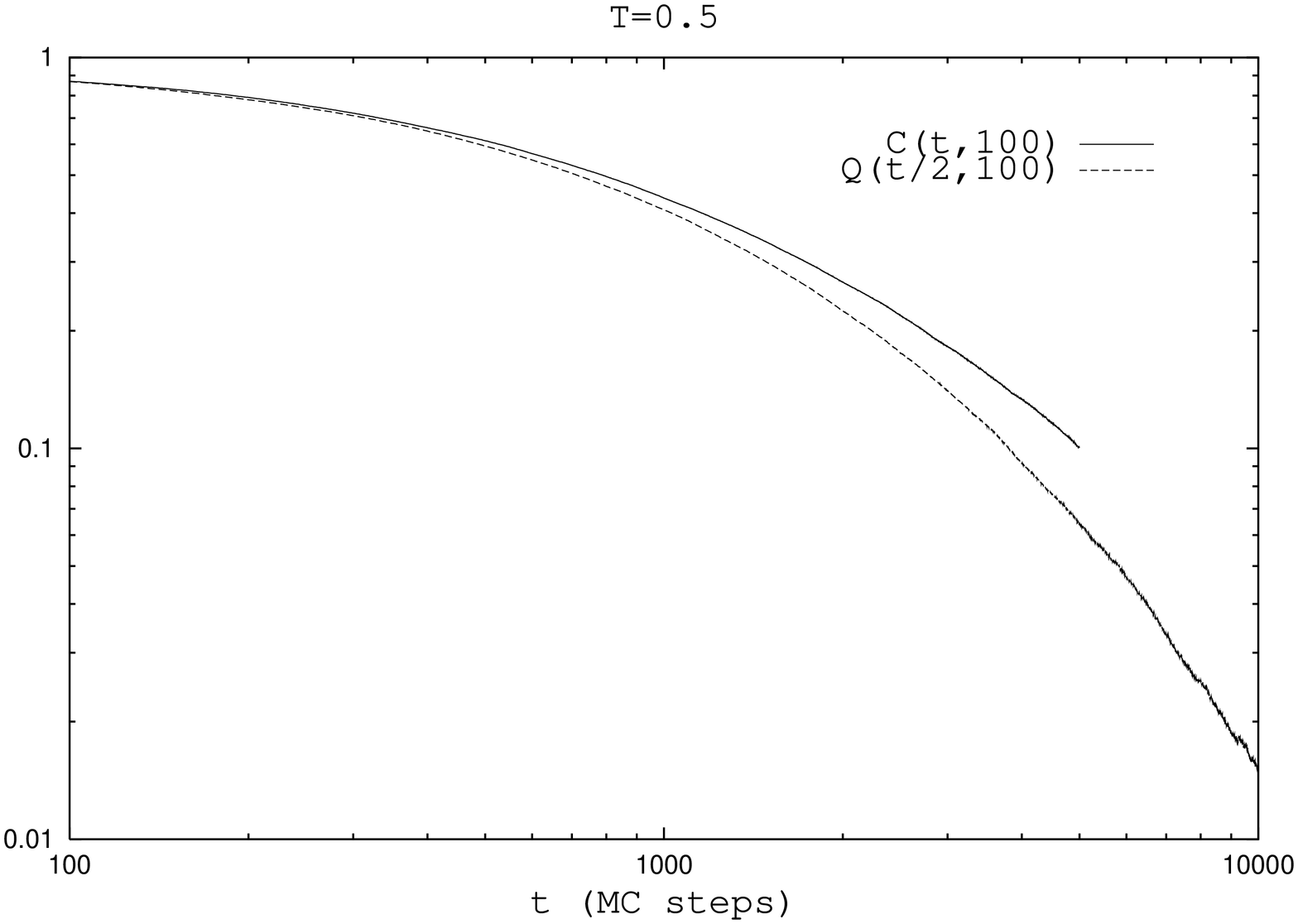}}
\subfigure[]{\includegraphics[scale=0.23,angle=-90]{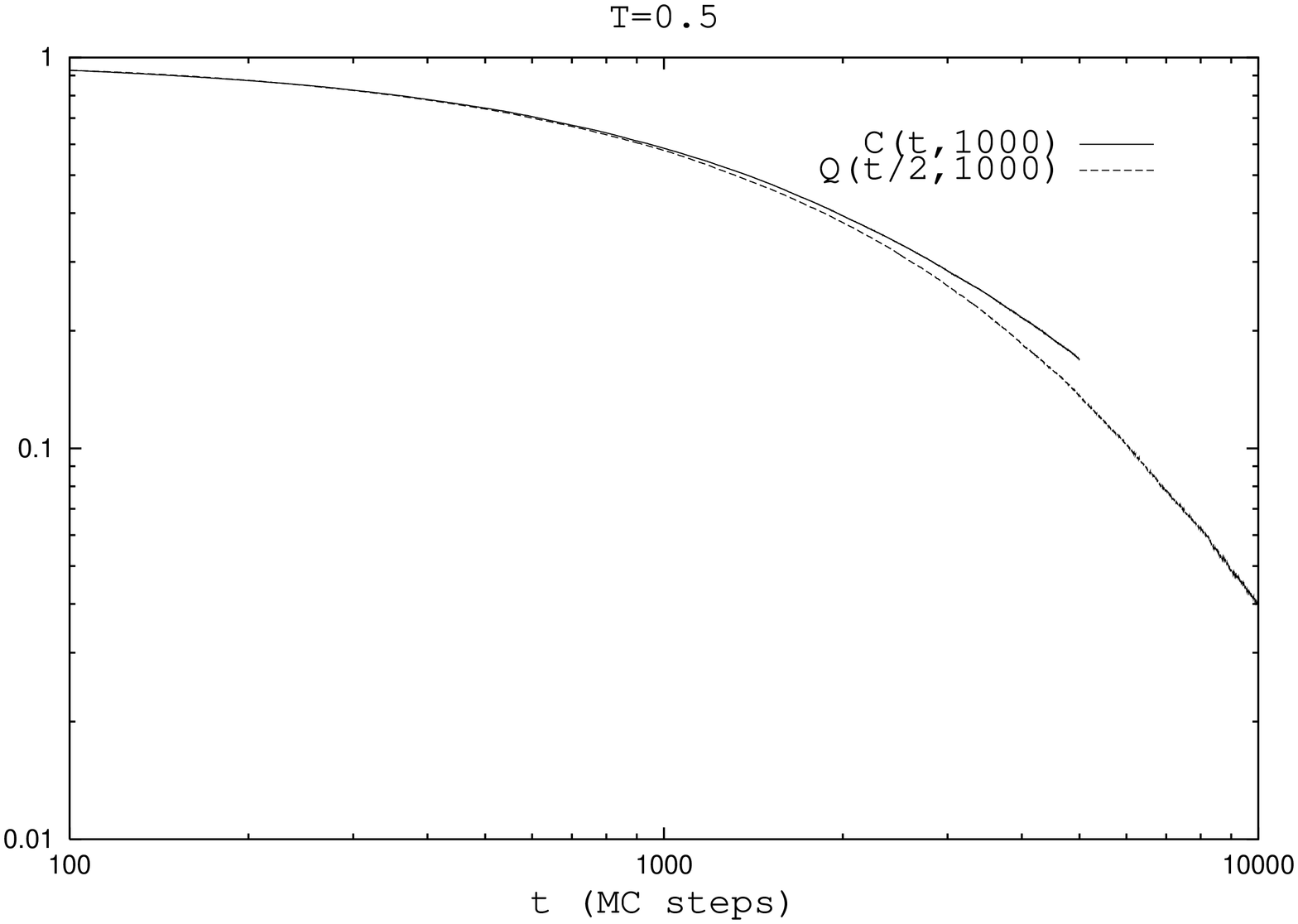}}
\end{center}
\caption{Plots of the functions $C(t,t_w)$ and $Q(t/2,t_w)$ for different
 temperatures and different $t_w$. In (a) and (b) we see how the scaling
 (\ref{non-glassy}) is satisfied for any $t_w$. In (c) and (d) we see
 that the relation of scaling is not satisfied at low temperatures.}
\label{QdeC}
\end{figure}
we can perform different kind of analysis. One of those is the
following. It's known that the system must satisfy the relation
\begin{equation}
C_{spin}(t_w,t)=Q(t_w,t/2)
\label{non-glassy}
\end{equation}
if the we are in an ordinary non-glassy
phase. As we
can see in fig.\ref{QdeC}a and fig.\ref{QdeC}b there are
temperatures\footnote{Remember that $T=0.8$ is already a temperature
  where, at waiting times we are considering, the energy
  auto-correlators are not $t_w$ independent} for which the behavior
of the $Q$ parameter with respect to the $C_{spin}$ two times
auto-correlator is what we expected, while there are lower
temperatures, like in fig.\ref{QdeC}c and fig.\ref{QdeC}d where the
relation (\ref{non-glassy}) is not satisfied at large enough times, for
the value of $t_w$ we simulate, although the discrepancy region is
moving to larger times as we increase $t_w$ as if it would like to
follow (\ref{non-glassy}) for large $t_w$ (see fig.\ref{convergeCdQ}).

\subsection{Conclusions}
At this point we have concluded the analysis of all our simulations
without any clear reason to believe that this 2 dimensional version of
Savvidy's gonihedric model has really glassy behavior; on the
contrary it seems to exhibit only very slow dynamics not related to real
glassiness of the model. 
This has to be further investigated to clear out what type of dynamics
is this model developing. 
Also $\kappa\ne 0$ has to be investigated although we think it will
follow the same kind of behavior that the $\kappa=0$ case. 

\section{The model of loops coupled to gravity}

There exist a way to extent this model to one coupled to gravity. To
couple it to gravity we will put our spin model in a random lattice
built from quadrangular pieces. In this way we will keep the behavior
of the loops and add the gravity degrees of freedom. In order to
make the matrix model solvable (or approximately solvable) we will make
the loops highly self-interacting, that means that the loops will
\begin{figure}[!ht]
\begin{center}
\includegraphics[scale=0.3]{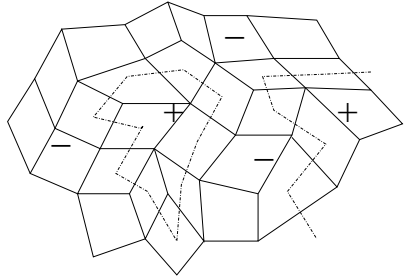}
\end{center}
\caption{Example of a random lattice with a gonihedric spin model on it}
\label{random lattice}
\end{figure}   
never cross themselves. This corresponds to the $\kappa\to\infty$
limit in the model we presented above. In fig.\ref{random lattice} we
can see an example of this kind of quadrangulations. 
From this picture we can extract the weights of each interaction term
in the matrix model that will represent the partition
function of our system. Let's see what those terms mean.

\begin{figure}[!ht]
\begin{center}
\includegraphics[scale=0.3]{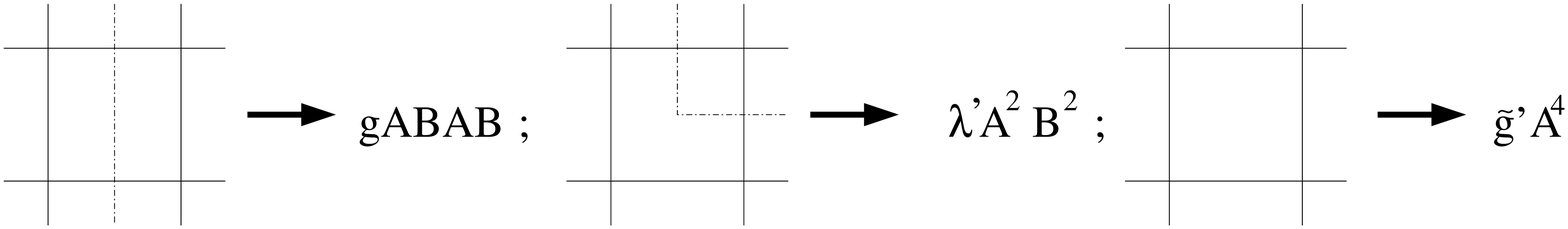}
\end{center}
\caption{correspondence between the loop pieces and the matrix
  interaction that are going to generate them.}
\label{intterms}
\end{figure}

First of all there will be the bulk term (in other words, plaquette
that is not crossed by any loop), then we have to consider a term were
a plaquette is crossed by one loop without bending through it and finally
the term were the loop crossing the plaquette bends in one direction or
the other. These three building blocks of the random lattices are
presented graphically in fig.\ref{intterms} with the corresponding
term of the matrix model that will generate them\footnote{To simplify the
  visual identification with the loop model we have not drawn the
  lines corresponding to the ``bulk'' propagator, i.e. to the
  A matrix propagator. The loop is generated with the
  B matrix propagator.}.
As we can see we are considering the most general case where all the
couplings are different but in our specific case we will impose the
condition $\tilde g=g$ due to the fact that a straight piece of loop
do not contribute with any amount of energy to the action, so the
coupling has to be equal to the bulk coupling.
In addition to those terms we have to put the kinetic term for the two
matrices, i.e. the quadratic terms $A^2$ and
$B^2$. So finally the matrix model that will represent our
loop model coupled to gravity will be,
\begin{equation}
e^Z=\int\!\!\!\int \dif A\dif B\ exp\big(-\Tr[A^2
+B^2+\frac{g}{N} A^4+\frac{\lambda^\prime}{N} A^2B^2+
\frac{\tilde g^\prime}{N} ABAB]\big) \nonumber
\end{equation}

Now that we have found the matrix model that will reproduce our loop
model coupled to gravity we only need to develop its solution. To our
knowledge this model has not been solved exactly. Although very
similar matrix models have been indeed solved \cite{Krietal} their
solution cannot be applied to our matrix model. 

\subsection{A partial solution to the matrix model}

Let's proceed then to the approximation to the solution. To this aim we
are going to rescale the $A$ matrix in the form $A
\to \sqrt{\frac{N}{g}}A$ so that the action will have the
following appearance.
\begin{equation}
S=\frac{N}{g}\Tr[A^2+A^4]+\Tr[B^2+
  \lambda A^2B^2+ \tilde
  gABAB] \nonumber
\end{equation}
where we have made some redefinitions like $\lambda^\prime / g\to\lambda$
and $\tilde g^\prime/g\to\tilde g$. Now we are going to pay attention to
the $B$--dependent part. As the action $S$ is quadratic in
$B$ we should be able to integrate out the $B$
matrix and find a one matrix model equivalent to the one we are using
now. 

\begin{figure}[!ht]
\begin{center}
\includegraphics[scale=0.3]{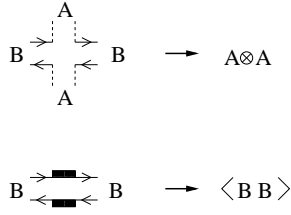}
\end{center}
\caption{Feynman rules for our matrix model}
\label{Feynmanrules}
\end{figure}

\begin{figure}[!ht]
\begin{center}
\includegraphics[scale=0.3]{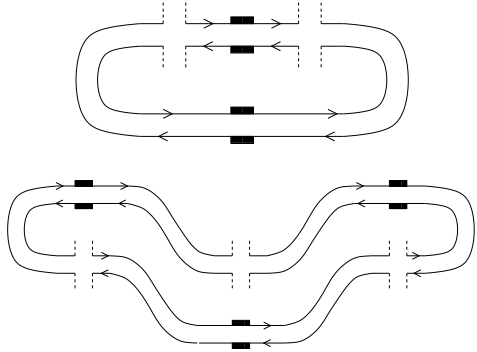}
\end{center}
\caption{two examples of connected diagrams that contribute to the
  integral. a non trivial contraction is shown in the second diagram,
  can be easily seen that is exactly equivalent to the trivial diagram.}
\label{diagrams}
\end{figure}

The integration we are facing now is the following
\begin{equation}\label{integral.1}
\int \dif B\
  exp\big(-\Tr[B(\mathbb{I}+\lambda A^2)B]- 
  \tilde g\ \Tr[ABAB]\big)
\end{equation}
Here we are going to interpret the first part of the action as the
free action (and rename $\mathbb{I}+\lambda A^2=M$) and
the second part as the interaction, so we can do perturbation theory and
re-sum all the diagrams at the end. 
But before we calculate diagrams we need to know the free propagator
of the $B$ matrix, and to reach this we add some external
currents and perform the quadratic integration\footnote{In appendix
  \ref{Bpropagator} is shown in detail how to make this calculation
  and find $\tilde\textrm{Z}(J,M)$.}. So finally
we find the propagator
\begin{equation}
\big<B_{ij}B_{kl}\big>=\tilde\textrm{Z}(0,M)
\bigg[\frac{\mathbb{I}\otimes\mathbb{I}}{\mathbb{I}\otimes M+
    M\otimes\mathbb{I}}\bigg]_{il;kj} 
\label{propagator}
\end{equation}
where $\tilde\textrm{Z}(0,M)$ a determinant coming from the
B integration.
Once we have found the propagator we can proceed with the diagrammatic. We
will only consider the connected diagrams. The Feynman rules for the
diagrams will be those shown in fig.\ref{Feynmanrules}
To order $n$ in the interaction term there will be also a factor
$-\tilde g/n!$ due to the expansion of the exponential, and a
combinatorial factor of $(n-1)!\ 2^{n-1}$ coming from the reordering of
the interaction terms ( in fig.\ref{diagrams} we can see the kind
of diagrams that will contribute). 
So finally all connected diagrams add up to
\begin{eqnarray}
&&\sum_{n=1}^\infty \frac{(-\tilde g)^n\ (n-1)!\ 2^{n-1}}{n!}
\Tr\bigg[(A\otimes A\ [\mathbb{I}\otimes M+
    M\otimes\mathbb{I}]^{-1})^n \bigg]\nonumber\\
&&\quad=\Tr\Bigg[\frac{1}{2}\sum_{n=1}^\infty \frac{(-2\tilde g)^n}{n}
  \bigg(A\otimes A[\mathbb{I}\otimes M+
    M\otimes\mathbb{I}]^{-1}\bigg)^n\Bigg]\nonumber\\
&&\quad=\frac{-1}{2}\Tr\Bigg[log\bigg(\mathbb{I}\otimes\mathbb{I}+
  \frac{2\tilde  gA\otimes A}{2
    \mathbb{I}\otimes\mathbb{I}+\lambda(\mathbb{I}\otimes A^2+
    A^2\otimes\mathbb{I})}\bigg) \Bigg]\nonumber
\end{eqnarray}
where we have recovered the explicit form of $M$ in terms of
$A$.
Thus exponentiating this last expression we recover all connected and
disconnected diagrams; i.e. the integral (\ref{integral.1}) we were
trying to calculate.
\begin{eqnarray}
\int \dif B\
  exp\big(-\Tr[B(\mathbb{I}+\lambda A^2)B]\ -
  \tilde g\ &&\rmspc \Tr[ABAB]\big)=\nonumber\\
=\tilde\textrm{Z}(0,M)\ exp\Bigg\{\frac{-1}{2}
  \Tr\Bigg[log\bigg(\mathbb{I}\otimes\mathbb{I}\ +\ &&\rmspc 
  \frac{2\tilde  gA\otimes A}{2
    \mathbb{I}\otimes\mathbb{I}+\lambda(\mathbb{I}\otimes A^2+
    A^2\otimes\mathbb{I})}\bigg)\Bigg]\Bigg\}\nonumber
\end{eqnarray}

So finally we have found an expression for the integration of the
$B$ matrix. In this expression $\tilde\textrm{Z}(0,M)=
\big[det(\mathbb{I}+\lambda A^2)\big]^{-\frac{N}{2}}$ that can
be included in the effective action as
$-\frac{N}{2}\Tr[log(\mathbb{I}+\lambda A^2)]$. This means that
we have rewritten our two matrix model in terms of a one matrix
model. So now we can diagonalize our remaining $A$ matrix and
integrate over the rotational degrees of freedom to leave our
partition function in the form
\begin{eqnarray}
e^Z&=&\int\!\!\!\int \dif A\dif B\ exp\big(-\Tr[A^2
+B^2+\tilde g^\prime A^4+\lambda A^2B^2+
g^\prime ABAB]\big)\nonumber\\
&=&\int \bigg(\prod_{i=1}^N\  \dif
a_i\bigg)\Delta^2(a)exp\Big\{-\frac{N}{\tilde
  g}\textrm{S}_{eff}\big(\{a\}\big)\Big\}\nonumber  
\end{eqnarray}
where $\Delta(a)=\prod_{i<j}(a_j-a_i)$ is the usual Van der Monde
determinant and $\textrm{S}_{eff}$ is the following
\begin{eqnarray}
\textrm{S}_{eff}\big(\{a\}\big)&=&\sum_{i=1}^N\Big(a_i^2+a_i^4+\frac{g}{2}
log(1+\lambda a_i^2)\Big)+\nonumber\\&&\sum_{i,j}^N\frac{\tilde
  g}{2N}log\bigg(1+\frac{2\tilde
  ga_ia_j}{2+\lambda(a_i^2+a_j^2)}\bigg) \nonumber
\end{eqnarray}

Since we are working in terms of eigenvalues we can use standard
procedures to try to solve the model. Using saddle point approximation
we arrive to a set of coupled equations that look quite difficult to
solve exactly. In the $N\to\infty$ limit those equations read
\begin{eqnarray}
&&\rmspc a+4a^3+g\Big[\frac{\lambda a}{1+\lambda a^2}+\int\dif b\rho (b)
  \frac{2\tilde gb\big(2+\lambda
  (b^2-a^2)\big)}{\big(2+\lambda(b^2+a^2)+2\tilde
  gba\big)\big(2+\lambda (b^2+a^2)\big)}\Big]\nonumber \\&&=2g\barint\dif
  b\frac{\rho(b)}{a-b}=
  -g(\omega(a+i\epsilon)+\omega(a-i\epsilon))\quad, \nonumber 
\end{eqnarray}
where as usual $\rho(a)=lim_{N\to\infty}\Big[\frac{1}{N}\Sigma_{i=1}^N
  \delta(a-a_i)\Big]$ and the resolvent $\omega(z)$ is defined to be
  equal to $\int\dif b\ \rho(b)/(b-z)$
This self-consistent equation has to be analyzed carefully to see
  whether there is any fixed point in the set of parameters that
  allowed us to pas to the continuum.

\subsection{Conclusions}

In the conclusions of this second part we may comment that, since the
fixed geometry model we have studied in the first part does not posses
any thermodynamical singularity, even for $\kappa\to\infty$, we would
naively expect the present matrix model not to exhibit any scaling
limit, but this issue deserves further analysis. Related to this
matrix model there are other matrix models that can be exactly solved
\cite{Krietal}. Although those other models have some critical
differences, their solutions may give some hints on how to exactly 
solve the model we are interested in. In fact some of the solved
models can be found as an special limit of ours. That could
be used as a test or a guide to find the solution.

\begin{center}
{\bf Aknowledgments}
\end{center}

We express our gratitude to the organizers of the workshop on Random
Geometry Krakow 2003 for a most enjoyable atmosphere and
hospitality. I would like to thank A.Dominguez and L. Tagliacozzo for
interesting discussions and encouragement. The research contained in
this talk is supported by the European Network EUROGRID and for a
CIRIT grant 2001FI-00387.

\appendix

\section{Finding the $B$ matrix propagator}\label{Bpropagator}
To find the propagator we begin with the free action and add an
external source. In our case this will lead to 
\begin{equation}
\Tr\big[BMB+JB\big]\nonumber
\end{equation}
Then, to reabsorb the external field $J$ into the
$B$ field we do a linear change of variables
$B\to\tilde B+C$ and quadratize the action. Doing
the inverse procedure we find
\begin{eqnarray}
&\tilde BM\tilde B-CMC
  =(B-C)M(B-C)-
  CMC& \nonumber\\
&=BMB-(CMB+
  BMC)=
  BMB-JB& \nonumber
\end{eqnarray}
So at the end the integral is
\begin{eqnarray}
\tilde\textrm{Z}(J,M)&=&\int\dif B\
exp\big\{-\Tr[BMB-JB]\big\}\nonumber\\
&=&\Big(det[M]\Big)^{-1}\
exp\big\{\Tr[CMC]\big\} 
\label{integral.2}
\end{eqnarray}
where the $C$ matrix can be determined from the equation
\begin{equation}
CM+MC=J 
\label{Cequation}
\end{equation}
Let's solve this equation to find the explicit form of the
$C$ matrix. To this aim let's choose the basis where the
$M$ matrix is diagonal. So if $\Omega $ is the matrix that
diagonalizes it
\begin{eqnarray}
&&M=\Omega \textrm{D}\Omega^\dagger\qquad \textrm{where}\qquad
\textrm{D}_{ij}=\delta_{ij}d_i\nonumber\\
&&J=\Omega J^\prime\Omega^\dagger\label{rotation}\\
&&C=\Omega C^\prime\Omega^\dagger\nonumber
\end{eqnarray}
The primed matrices correspond to the non-primed ones after the rotation.

So now we will continue solving the equation (\ref{Cequation}) by
writing it in the $M$--diagonal form,
\begin{eqnarray}
(C^\prime\textrm{D}+\textrm{D}C^\prime)_{ij}&=&
  C^\prime_{ij}d_j+d_iC^\prime_{ij}=
  J^\prime_{ij}\nonumber\\ 
  C^\prime_{ij}&=&\frac{J^\prime_{ij}}{d_i+d_j}\nonumber
\end{eqnarray}
Now introduce it in eq.(\ref{integral.2}) and invert
eq.(\ref{rotation}) to find
\begin{eqnarray}
&&\rmspc\textrm{Z}(J,M)=\Big(det[M]\Big)^{-1}\
\exp\Bigg\{ \sum^N_{i,j=1} \bigg[\frac{J_{ij}^\prime}{d_i+d_j}
  d_j\frac{J_{ji}^\prime}{d_j+d_i}\bigg]\Bigg\}\nonumber\\ 
&&=\Big(det[M]\Big)^{-1}
\exp\Bigg\{ \sum^N_{i,j=1} \frac{d_j}{(d_i+d_j)^2}
\big[\Omega_{ik}^\dagger J_{kl}\Omega_{lj}\big]
\big[\Omega_{jm}^\dagger J_{mn}\Omega_{ni}\big]\Bigg\} \nonumber
\end{eqnarray} 
That is the last expression we'll write for
$\textrm{Z}(J,M)$. From here we can deduce all
necessary correlators, like the one we are looking for; the propagator
$\big<B_{ij}B_{kl}\big>=
\delta\textrm{Z}(J,M)/
\delta J_{ji}\delta J_{lk}\mid_{J=0}$.
Calculating those variations and rotating back the $D$ matrices we
find
\begin{eqnarray}
&&\rmspc\textrm{Z}(0,M)\Bigg[\sum_{m,n=1}^N\frac{d_n}{d_m^2+d_n^2}\Big\{
  \big(\Omega^\dagger_{ml}\Omega_{kn}\big)
  \big(\Omega^\dagger_{nj}\Omega_{im}\big)
+ \big(\Omega^\dagger_{mj}\Omega_{in}\big)
  \big(\Omega^\dagger_{nl}\Omega_{km}\big)\Big\}\Bigg] \nonumber\\
&&\quad=\textrm{Z}(0,M)\Bigg[\bigg[\frac{M\otimes\mathbb{I}}
    {[\mathbb{I}\otimes M+M\otimes\mathbb{I}]^2}\bigg]
  +\bigg[\frac{\mathbb{I}\otimes M}{[\mathbb{I}\otimes M
	+M\otimes\mathbb{I}]^2}\bigg]\Bigg]_{il;kj} \nonumber\\
&&\quad=\textrm{Z}(0,M)\Bigg[\frac{\mathbb{I}\otimes\mathbb{I}}
  {\mathbb{I}\otimes M +M\otimes\mathbb{I}}\Bigg]_{il;kj}
\end{eqnarray}


\begin{thebibliography}{99}
\bibitem{Amba-sukiasian} R.V: Ambartzumian, G.S. Sukasian, G.K. Savvidy and K.G. Savvidy, Phys. Lett. {\bf B275} (1992) 99;\\
G.K. Savvidy and K.G. Savvidy, Int. J. Mod. Phys. {\bf A8} (1993) 3393;\\ G.K. Savvidy and K.G. Savvidy, Mod. Phys. Lett.
{\bf A8} (1993) 2963.
\bibitem{savvidy} G.K. Savvidy and F.J. Wegner, Nucl. Phys. {\bf B413} (1994) 605;\\
G.K. Savvidy and K.G. Savvidy, Phys. Lett. {\bf B324} (1994) 72;\\
G.K. Savvidy, K.G. Savvidy and F.J. Wegner, Nucl. Phys. {\bf B443}
(1995) 565;\\ J. Ambjorn, G. Koutsoumbas and G.K. Savvidy, Europhys.Lett. {\bf 46} (1999) 319 (cond-mat/9810271).
\bibitem{kuts} G. Koutsoumbas, G.K. Savvidy and K.G. Savvidy,
  Phys.Lett. {\bf B410} (1997) 241 
(hep-th/9706173). 
\bibitem{baig} M. Baig, D.Espriu, D. Johnston and R.P.K.C. Malmini,
J.Phys. {\bf A30} (1997) 7695 (hep-lat/9703008). 
\bibitem{esbaijo} D.Espriu, M. Baig, D.A.Johnston, R.K.P.C.Malmini,
J.Phys. {\bf A30} (1997) 405 (hep-lat/9607002).
\bibitem{jonmal} D.A. Johnston and R.P.K. Malmini, Phys.Lett. {\bf
  B378} (1996) 87, (hep-lat/9508026). 
\bibitem{eslipjo} A.Lipowski, D.Johnston and D.Espriu, Phys.Rev. {\bf
  E62} (2000) 3404  
(cond-mat/0004466). 
\bibitem{swift} M.R. Swift, H. Bokil, R.D.M. Travasso and A.J. Bray,
  Phys.Rev. {\bf B62} 11494 (2000) (cond-mat/0003384). 
\bibitem{lip} A. Lipowski, J. Phys. A: Math. Gen. {\bf 30} (1997) 7365.
\bibitem{lipjo} A. Lipowski and D. Johnston, {\it Glassy transition
  and metastability in four-spin Ising model} (cond-mat/9812098).
\bibitem{lipjo_}A. Lipowski and D. Johnston, Phys.Rev. {\bf E61}, 6375
(2000)(cond-mat/9910370).\\ 
A. Lipowski and D. Johnston, Phys.Rev. {\bf E64}, 041605
(2001)(cond-mat/0105602). 
\bibitem{DEJP}P. Dimopoulos, D. Espriu, E. Jane, A. Prats. Phys.Rev.E
  66:056112, 2002. [COND-MAT 0204403]  
\bibitem{sav} G.K.Savvidy, {\it The system with exponentially
  degenerate vacuum state}
  (cond-mat/0003220). 
\bibitem{kutsanis} G. Koutsoumbas and G.K. Savvidy, {\it
  Three--dimensional gonihedric spin system} 
(cond-mat/0111590). 
\bibitem{mezard} A. Barrat, R. Burioni and M. Mezard,  J. Phys. {\bf
  A29} (1996) 1311 (cond-mat/9509142). 
\bibitem{Krietal} L. Chekhov, C. Kristjansen.
    Nucl.Phys.B479:683-696,1996.
    [HEP-TH 9605013]\\
 B. Eynard, C. Kristjansen.
    Nucl.Phys.B516:529-542,1998.
    [COND-MAT 9710199]\\
 Ivan K. Kostov.
    Phys.Lett.B549:245-252,2002.
    [HEP-TH 0005190]






\end{thebibliography}
\end{document}